\documentclass[preprint,12pt]{aastex}

\usepackage[footnotesize]{caption}

\shorttitle{\emph{SMA} observations of the G240.31+0.07 Outflow}
\shortauthors{Qiu et al.}
\begin{document}

\title{Submillimeter Array Observations of the Molecular Outflow
in High-mass Star-forming Region G240.31+0.07}

\author{Keping Qiu\altaffilmark{1,2}, Qizhou Zhang\altaffilmark{2},
Jingwen Wu\altaffilmark{2}, Huei-Ru Chen\altaffilmark{3,4}}




\altaffiltext{1}{Department of Astronomy, Nanjing University,
Nanjing, China} \altaffiltext{2}{Harvard-Smithsonian Center for
Astrophysics, Cambridge, MA} \altaffiltext{3}{Institute of
Astronomy and Department of Physics, National Tsing Hua
University, Hsinchu, Taiwan} \altaffiltext{4}{Institute of
Astronomy and Astrophysics, Academia Sinica, Taiwan}

\begin{abstract}
We present Submillimeter Array observations toward the $10^{4.7}$
L$_{\odot}$ star-forming region G240.31+0.07, in the $J=2$--1
transition of $^{12}$CO and $^{13}$CO and at 1.3 mm continuum, as
well as the $^{12}$CO and $^{13}$CO observations from the Caltech
Submillimeter Observatory to recover the extended emission
filtered out by the interferometer. Maps of the $^{12}$CO and
$^{13}$CO emission show a bipolar, wide-angle, quasi-parabolic
molecular outflow, roughly coincident with an IR nebula revealed
by the \emph{Spitzer} 3.6 and 4.5 $\mu$m emission. The outflow has
$\sim98$ M$_{\odot}$ molecular gas, making it one of the most
massive molecular outflows known, and resulting in a very high
mass-loss rate of $4.1\times10^{-3}$ M$_{\odot}$yr$^{-1}$ over a
dynamical timescale of $2.4\times10^4$ yr. The 1.3 mm continuum
observations with a $4''\times3''$ beam reveal a flattened dusty
envelope of $\sim150$ M$_{\odot}$, which is further resolved with
a $1.2''\times1''$ beam into three dense cores with a total mass
of $\sim40$ M$_{\odot}$. The central mm core, showing evidence of
active star formation, approximately coincides with the geometric
center of the bipolar outflow thus most likely harbors the
powering source of the outflow. Overall our observations provide
the best case to date of a well-defined wide-angle molecular
outflow in a $>10^4$ L$_{\odot}$ star-forming region. The outflow
is morphologically and kinematically similar to low-mass
protostellar outflows but has two to three orders of magnitude
greater mass, momentum, and energy, and is apparently driven by an
underlying wide-angle wind, hence further supports that high-mass
stars up to late-O types, even in a crowded clustering
environment, can form as a scaled-up version of low-mass star
formation.

\end{abstract}


\section{Introduction}
Single-dish surveys have shown that molecular outflows are
ubiquitous in high-mass star forming regions \citep{Shepherd96,
Zhang01, Beuther02a}. According to the standard paradigm of
low-mass star formation this provides a statistical argument that
high-mass stars may form via accretion processes. However, given
the typical large distances ($>1$ kpc) and clustering environment
of massive star formation, single-dish observations of $>10''$
resolutions cannot resolve the detailed structures of massive
molecular outflows or identify their driving sources. Recently
there have been an increasing number of interferometric studies
which provide new insights into the morphology and kinematics of
massive molecular outflows \citep[e.g.,][and references
therein]{Qiu07}, but the statistics is still poor and many basic
questions, e.g., how massive molecular outflows systematically
differ from their low-mass counterparts, how they are driven, and
how they affect their surroundings and host clouds, remain open.
To address these questions, which are essential to understand
massive star formation, a large set of high-angular-resolution
observations covering a wide range of luminosity and evolutionary
stages is needed.

Toward these goals we study G240.31+0.07 (hereafter G240), an
active high-mass star-forming region with a far-infrared
luminosity of $10^{4.7}$ L$_{\odot}$ at a distance of 6.4 kpc
\citep{Macleod98}. It harbors an ultra-compact (UC) H{\scriptsize
II} region and is associated with H$_2$O and OH masers
\citep{Hughes93, Caswell97, Macleod98, Migenes99}. \citet{Kumar03}
reported a parsec-scale velocity gradient in C$^{18}$O (2--1) with
a $33''$ beam and the detection of near-infrared H$_2$ emission in
this region. \citet{Chen07} presented an interferometric study at
654 GHz of the central core and resolved the dust continuum
emission into two clumps both associated with C$^{18}$O (6-5)
emission. In the meantime, molecular outflow gas in this region
has been mapped using single-dish telescopes: \citet{Shepherd96}
reported a CO (1--0) bipolar outflow with a $60''$ beam;
\citet{Hunter97} mapped the CO (3--2) emission with a $20''$ beam
and proposed two overlapping outflows both in the
northwest-southeast orientation. However, limited by low angular
resolutions, neither of the two single-dish studies provided a
detailed analysis of the outflow morphology and kinematics.
High-angular-resolution observations are crucial to unveil the
outflow nature of this luminous source.

Here we present a $^{12}$CO, $^{13}$CO (2--1) study of the G240
outflow, using the Submillimeter Array\footnote[5]{The
Submillimeter Array is a joint project between the Smithsonian
Astrophysical Observatory and the Academia Sinica Institute of
Astronomy and Astrophysics and is funded by the Smithsonian
Institution and the Academia Sinica.} (SMA) observations.
Complementary short spacings are obtained from the Caltech
Submillimeter Observatory\footnote[6]{The Caltech Submillimeter
Observatory is supported by the NSF grant AST-0229008} (CSO) 10.4
m telescope. The archival \emph{Spitzer} mid-infrared imaging is
also presented for comparison. The 1.3 mm continuum observations
are used to pinpoint the outflow powering source.

\section{Observations}
The SMA observations were carried out on 2008 February 25 with 8
antennas in the compact configuration and on 2008 February 16 with
7 antennas in the extended configuration. The $2\times2$GHz
correlator was configured to cover $^{12}$CO (2--1) in the upper
sideband and $^{13}$CO (2--1) in the lower sideband. And the
continuum was constructed from line-free channels. Lines of
$^{12}$CO, $^{13}$CO (2--1) and the 1.3 mm continuum, which are
relevant to the outflow study, are presented in this work. Many
other lines covered in the same setup will be presented in a
future paper. To ensure the coverage of the entire outflow we
observed two fields centered at (R.A., decl.)$_{\mathrm
J2000}=(07^{\mathrm h}44^{\mathrm m}52.49^{\mathrm s},
-24^{\mathrm d}07^{\mathrm m}52.1^{\mathrm s})$ and (R.A.,
decl.)$_{\mathrm J2000}=(07^{\mathrm h}44^{\mathrm
m}51.07^{\mathrm s}, -24^{\mathrm d}07^{\mathrm m}34.9^{\mathrm
s})$, respectively. We used Titan as the primary flux calibrator
and 3c273 as the bandpass calibrator. The time dependent gain was
monitored by observing quasars J0730-116 and J0826-225 every 23
mins. Visibilities were calibrated using the IDL MIR package and
then output to MIRIAD for imaging. With natural weighting the
synthesized beams in the compact and extended configurations are
about $4''\times3''$ and $1.2''\times1''$, respectively.

The shortest baseline in our SMA observations is 16.5 m,
corresponding to a spatial scale of $20''$ for observations at 225
GHz. Thus spatial structures more extended than $20''$ were not
sampled in the SMA observations. This spatial filtering can
significantly affect the $^{12}$CO, $^{13}$CO (2--1) maps at
velocities close to the cloud velocity. To recover the missing
short spacing information we observed the $^{12}$CO, $^{13}$CO
(2--1) emission with the CSO 10.4 m telescope on 2008 February 12.
During the observation the weather condition was excellent for 1
mm waveband with ${\tau}_{225{\mathrm {GHz}}}\sim0.08$. The
observations were carried out in the on-the-fly mode centered on
(R.A., decl.)$_{\mathrm J2000}=(07^{\mathrm h}44^{\mathrm
m}52.1^{\mathrm s}, -24^{\mathrm d}07^{\mathrm m}49^{\mathrm s})$.
We obtained a $15\times15$ grid in $^{12}$CO and a $13\times13$
grid in $^{13}$CO, with an integration of about 5 s on each cell.
The $10''$ cell spacing is about $\frac{1}{3}$ of the CSO beam,
which is $\sim32.5''$ in $^{12}$CO (2--1). By observing Mars and
Saturn we derived a beam efficiency of $0.58\pm0.10$. The
spectrometer used has 1024 channels throughout the 50 MHz
bandwidth, resulting in a spectral resolution of 0.0488 MHz (about
0.065 km s$^{-1}$) per channel. The final maps were smoothed to 2
km s$^{-1}$ per channel. The data were reduced using the standard
CLASS package. We combined the SMA compact and CSO $^{12}$CO,
$^{13}$CO data in MIRIAD following a procedure outlined in
\citet{Zhang95}.

The deep \emph{Spitzer} IRAC images were obtained from the
\emph{Spitzer} archive (\dataset[ads/sa.spitzer#0014782464]{PID:
20499}). We found no obvious artifacts from a visual inspection,
thus adopted s14.0 Post Basic Calibrated Data products provided by
the \emph{Spitzer} Science Center.

\section{Results}
\subsection{mm Continuum Emission}
\begin{figure}
\epsscale{1.} \plotone{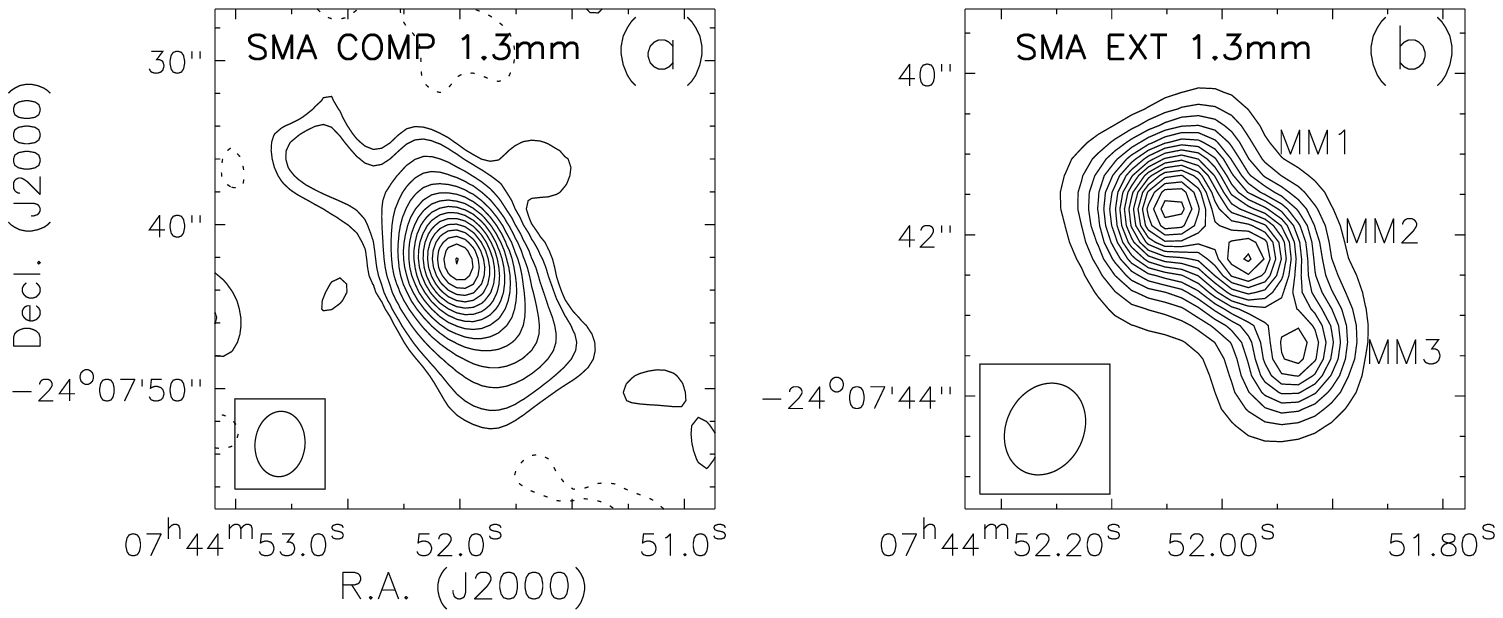} \caption{1.3 mm continuum emission:
(a) observed with the SMA compact array, contouring at
$3\times$(-1, 1, 2, 4, 7, 11, 16, 22, 29, 37, 46, 56, 67, 79, 92,
106) mJy beam$^{-1}$, where the r.m.s. noise level ($1\sigma$) is
about 1 mJy beam$^{-1}$; (b) observed with the SMA extended array,
contouring from 3.9 ($3\sigma$) to 54.6 mJy beam$^{-1}$ at steps
of 3.9 mJy beam$^{-1}$, and the three mm continuum peaks are
denoted as ``MM1'', ``MM2'', ``MM3'' from north to south. In the
field all the negative contour levels are higher than $-3\sigma$
thus not shown. The corresponding synthesized beam is displayed in
the lower left of each panel. \label{continuum}}
\end{figure}

In Figure \ref{continuum}a, the 1.3 mm continuum emission observed
with the SMA compact array shows an elongated envelope with its
major axis in the northeast-southwest direction, as well as a
minor peak to the northeast. At a higher angular resolution of
$1.2''\times1''$ the central part of the structure is resolved
into three dusty cores, namely MM1, MM2, and MM3 from north to
south (Figure \ref{continuum}b). Assuming optically thin dust
emission, the dust mass can be estimated according to
$$M_{dust}=\frac{F_{\nu}D^2}{B_{\nu}(T_{dust}){\kappa}_{\nu}}\,,$$ where
$M_{dust}$ is the dust mass, $F_{\nu}$ the continuum flux at
frequency $\nu$, $D$ the source distance, $B_{\nu}(T_{dust})$ the
Planck function at dust temperature $T_{dust}$, and
${\kappa}_{\nu}=10({\nu}/1.2\,{\mathrm {THz}})^{\beta}$ in
cm$^2$g$^{-1}$ the dust opacity \citep{Hildebrand83}. Based on SMA
observations toward this region \citet{Chen07} derived a gas
rotational temperature of $96$ K by comparing the brightness of
two H$_2$CO transitions, and derived an opacity index $\beta$ of
$1.5$ by comparing flux densities at 223 and 654 GHz. We then
adopt $T_{dust}=100$ K, $\beta=1.5$, and a gas-to-dust ratio of
100, to estimate the mass of the larger envelope and the combined
mass of the three dusty cores, which amount to 150 M$_{\odot}$ and
40 M$_{\odot}$, respectively. The uncertainty of this estimate
mainly comes from the determinations of $\beta$ and $T_{dust}$;
for example, deviations of $\pm0.3$ in $\beta$ or $\pm50$ K in
$T_{dust}$ will induce $\sim50$\% uncertainty in the mass
estimates.

\subsection{Bipolar Molecular Outflow}

\subsubsection{Outflow morphology and kinematics} 
\begin{figure}
\epsscale{1.} \plotone{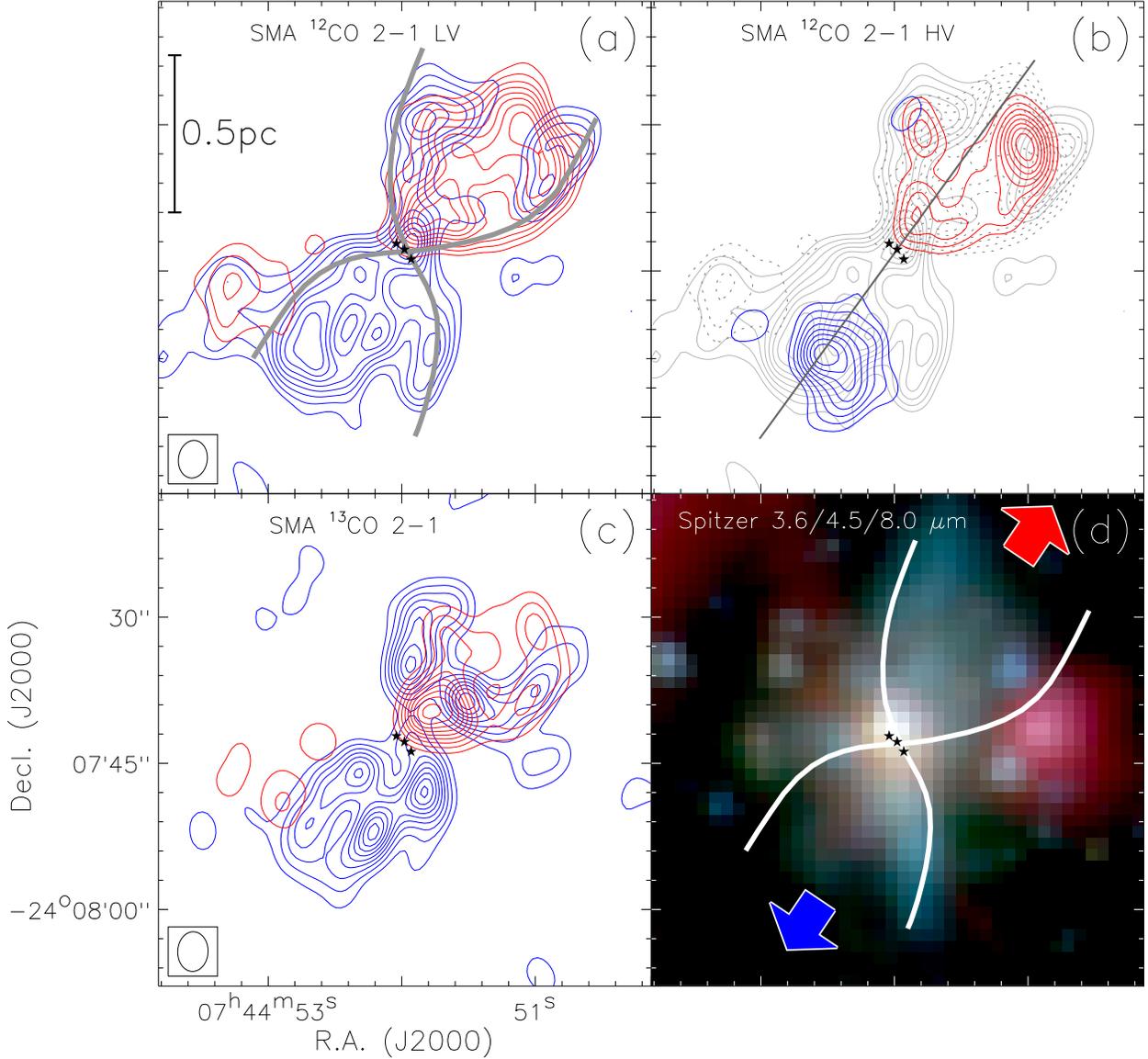} \caption{(a) low-velocity $^{12}$CO
2--1 emission, integrated from 58 to 64 km s$^{-1}$ for the
blueshifted lobe ({\it blue}) and from 74 to 80 km s$^{-1}$ for
the redshifted lobe ({\it red}), with contours starting from 20\%
and continuing at steps of 10\% of the peak emission, and two gray
curves bisecting each other at the central mm peak outlines the
shell of a bipolar wide-angle outflow, which is the focus of this
paper; (b) high-velocity $^{12}$CO 2--1 emission, integrated from
42 to 56 km s$^{-1}$ for the blueshifted lobe ({\it blue}) and
from 82 to 94 km s$^{-1}$ for the redshifted lobe ({\it red}),
with contours starting from 20\% and continuing at steps of 10\%
of the peak emission, the gray contours show low-velocity
$^{12}$CO emission as shown in panel (a), and a straight gray line
delineates the cut for a PV plot shown in Figure \ref{pv}; (c)
$^{13}$CO 2--1 emission, integrated from 58 to 62 km s$^{-1}$ for
the blueshifted lobe ({\it blue}) and from 74 to 78 km s$^{-1}$
for the redshifted lobe ({\it red}), with contours starting from
15\% and continuing at steps of 10\% of the peak emission; (d)
\emph{Spitzer} IRAC color composite image with the 3.6, 4.5, and
8.0 $\mu$m emission coded in blue, greed, and red, respectively,
the two curves are the same as shown in panel (a) to denote the
molecular outflow shell, and the blue and red arrows mark the
orientations of the blue- and redshfited lobes of the outflow,
respectively. In (a) and (c) the corresponding synthesized beam is
shown as an ellipse in the lower left. In each panel and
hereafter, the central stars denote the three mm continuum peaks.
\label{flow_int}}
\end{figure}

Figure
\ref{flow_int} shows the integrated $^{12}$CO, $^{13}$CO (2--1)
emission observed with the SMA, as well as the \emph{Spitzer} IRAC
imaging for comparison. Because the interferometric filtering to
extended emission helps to distinguish the outflow from ambient
gas, we present the SMA observations without merging with the
single-dish data. The cloud velocity ($v_{cloud}$) with respect to
the local standard of rest is about 67.5 km s$^{-1}$, which is
adopted from the CSO C$^{18}$O (2--1) observations by
\citet{Kumar03} and consistent with our SMA C$^{18}$O (2--1)
observations. The low-velocity channels, i.e., 58--64 km s$^{-1}$
for the blueshifted lobe and 74--80 km s$^{-1}$ for the redshifted
lobe in Figure \ref{flow_int}a, are chosen to highlight wide-angle
structures and to avoid contamination from diffuse gas in the
surrounding, and the high-velocity channels (Figure
\ref{flow_int}b) are those in the outer line wings. The
low-velocity $^{12}$CO emission clearly reveals a
northwest-southeast (NW-SE) bipolar outflow centered at the group
of the mm peaks and extending about 1.3 pc from end to end (Figure
\ref{flow_int}a). The outflow shell highlighted by
limb-brightening outlines a quasi-parabolic structure with a wide
opening angle. Both the spatial separation of the two lobes of
such a wide-angle outflow and the partial overlap of blue- and
redshifted emission in the NW lobe suggest that the flow axis is
moderately inclined to the plane of sky. We suggest the molecular
outflow seen in the $^{12}$CO emission as ambient gas swept-up by
an underlying wide-angle wind (see \S \ref{comparison}). In this
context, the NW wind is overall moving away from us, sweeping up
ambient gas to create the redshifted $^{12}$CO outflow, but could
be partly moving toward us in its near side, causing the
blueshifted $^{12}$CO emission. The patchy appearance of the
blueshifted emission might be attributed to the localized density
structure of the medium. The SE counterpart of the wind may be
expanding into a medium which is less dense in the far-side and
would not create appreciable redshifted $^{12}$CO emission.
Immediately to the southeast of the bipolar outflow, there is a
minor extension with overlapping red- and blueshifted emission,
which might arise from another outflow. Meanwhile, the
high-velocity $^{12}$CO emission coincides with the inner part of
the low-velocity bipolar outflow, with the bulk blueshifted
emission revealing a clump in the middle of the low-velocity SE
lobe and the redshifted emission a conical structure narrower than
the low-velocity NW lobe (Figure \ref{flow_int}b). The bipolar
outflow is also detected in the $^{13}$CO emission with a
morphology similar to but slightly less extended than the
$^{12}$CO outflow (Figure \ref{flow_int}c).

\emph{Spitzer} IRAC images often reveal outflow cavities through
scatted light in the 3.6 $\mu$m band and molecular outflow driving
agents (e.g., underlying jets or winds) through shocked H$_2$
emission in the 4.5 $\mu$m band \citep{Qiu08}. In Figure
\ref{flow_int}d, the IRAC three-color composite image of the G240
central area yields a NW-SE nebula in the 3.6 and 4.5 $\mu$m
emission. The north to south shell of the nebula appears prominent
in the 3.6 $\mu$m band ({\it blue}) and coincides well with the
north to south edge of the $^{12}$CO outflow, suggesting its
emission being dominated by scattered light from the outflow
cavity wall, whereas the nebula structure coincident with the main
NW-SE $^{12}$CO lobe shows diffuse 3.6 and 4.5 $\mu$m emission,
which might be dominated by scattered light and shocked H$_2$
emission.

\begin{figure}
\epsscale{.7}\plotone{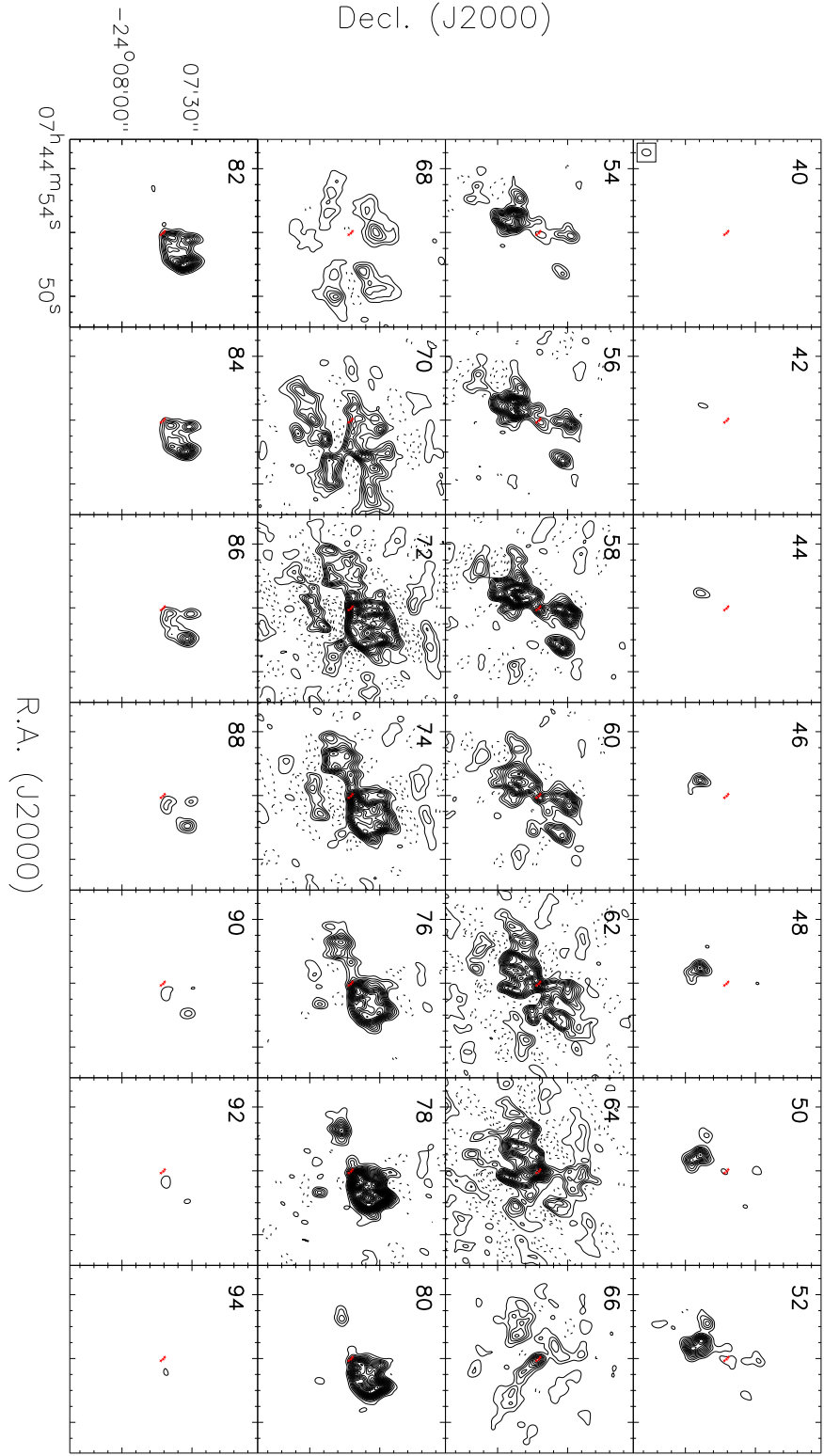} \caption{$^{12}$CO (2--1) channel
maps from the SMA compact array observations. Solid/dashed
contours start from and continue at steps of 0.5/-0.5 Jy
(+/-$10\sigma$) for channels of 40--58 km s$^{-1}$ and of 78--94
km s$^{-1}$, and 1.0/-1.0 Jy for channels of 60--76 km s$^{-1}$.
Hereafter the central velocity of each channel is indicated in the
upper left of each panel, and the synthesized beam is shown as an
ellipse in the lower left of the first panel. \label{sma12ch}}
\end{figure}

\begin{figure}
\epsscale{1.}\plotone{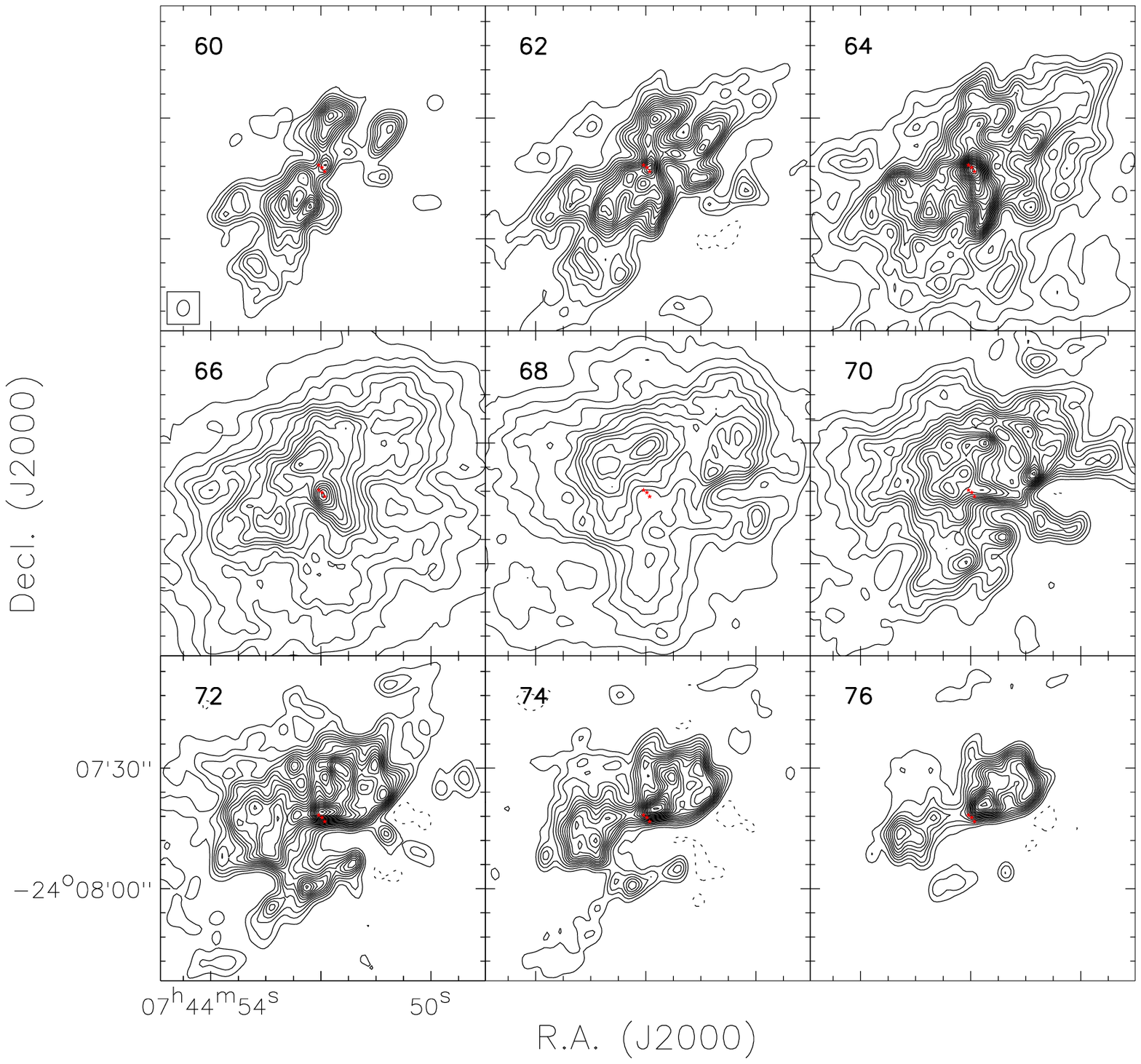} \caption{Combined SMA and CSO
$^{12}$CO (2--1) channel maps of 60--76 km s$^{-1}$, with
solid/dashed contours starting from and continuing at steps of
1.0/-1.0 Jy. \label{smacso12ch}}
\end{figure}

\begin{figure}
\epsscale{1.}\plotone{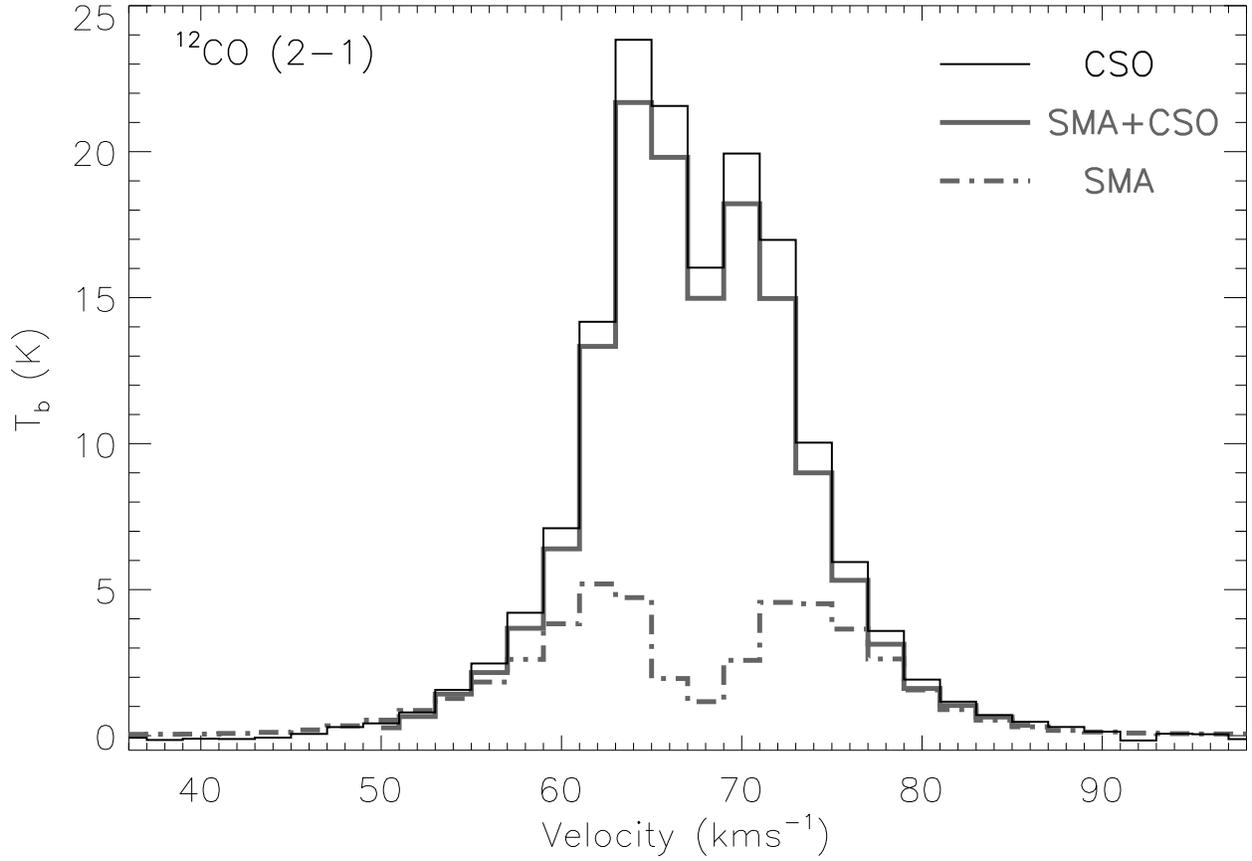} \caption{The $^{12}$CO (2--1)
spectra at the CSO observational center, (R.A., decl.)$_{\mathrm
J2000}=(07^{\mathrm h}44^{\mathrm m}52.1^{\mathrm s}, -24^{\mathrm
d}07^{\mathrm m}49^{\mathrm s})$: the solid black line shows the
CSO spectrum; the solid gray line shows the spectrum constructed
from the combined SMA and CSO data and convolved with the CSO
beam; the dash-dotted gray line shows the spectrum obtained with
the SMA compact array and convolved with the CSO beam.
\label{spec}}
\end{figure}

Figures \ref{sma12ch} and \ref{smacso12ch} show channel maps of
the $^{12}$CO emission from the SMA observations alone and from
the combined SMA and CSO data at low velocities, respectively. It
is immediately evident that for channels around $v_{cloud}$ the
combined data effectively recover the extended emission filtered
out by the interferometer and significantly improve the side-lobes
of the interferometric observations. In particular the diffuse gas
emissions at 66 and 68 km s$^{-1}$ are mostly filtered out by the
interferometer and recovered with the combined data. From a
comparison among the spectra constructed from the CSO, SMA, and
the combined data (Figure \ref{spec}), the SMA compact array
observations recover less than $10\%$ of the CSO flux at the line
center, and about $80\%$ of the CSO flux at the line wing of 54
and 80 km s$^{-1}$, whereas the combined SMA and CSO data recover
more than 85\% of the CSO flux all the way from 54 to 80 km
s$^{-1}$; at the outer line wing of $\leq$52 km s$^{-1}$ and
$\geq$82 km s$^{-1}$, there is little missing flux in the SMA
observations.

Outflow features emerge at 64 and 70 km s$^{-1}$; the SE
wide-angle shell at 64 km s$^{-1}$ appears less prominent in
Figure \ref{smacso12ch} than that in Figure \ref{sma12ch} due to
the confusion from diffuse gas. Emissions from outflow structures
appear predominant in channels of ${\leq}62$ km s$^{-1}$ and
${\geq}72$ km s$^{-1}$. The two lobes of the wide-angle outflow
accord with each other in orientation and have similar opening
angles at the base. They both appear more collimated at higher
velocities: the NW lobe has a fan-shaped morphology at low
velocities (most prominent at 72 and 74 km s$^{-1}$) and a
narrower, {\emph V}-shaped appearance at high velocities (most
prominent at 84 to 88 km s$^{-1}$); the SE lobe shows the
wide-angle shell at low velocities (most prominent at 62 and 64 km
s$^{-1}$) and a clump away from the mm peaks at high velocities
(most prominent at 48 to 52 km s$^{-1}$). Again the difference in
detailed velocity structures of the two lobes could be at least in
part attributed to the localized density structure of the
environment within which the underlying wind is expanding. As seen
in the integrated emission, a possible secondary outflow to the
southeast of the mm peaks can be found in both blue- and
redshifted channels.

\begin{figure}
\epsscale{.7}\plotone{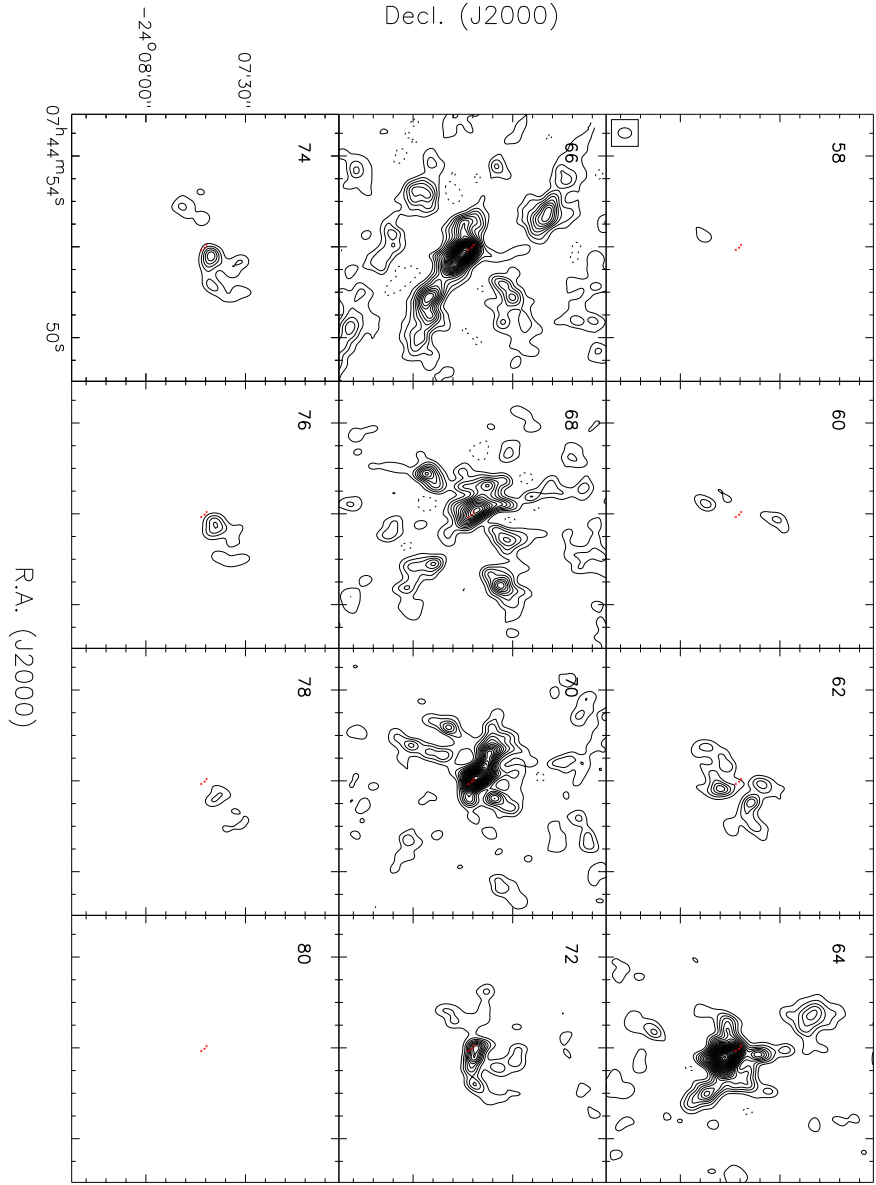} \caption{$^{13}$CO (2--1) channel
maps of 58--80 km s$^{-1}$, from the SMA compact array
observations. Solid/dashed contours start from and continue at
steps of 0.3/-0.3 Jy (+/-$6\sigma$). \label{sma13ch}}
\end{figure}

\begin{figure}
\epsscale{.8}\plotone{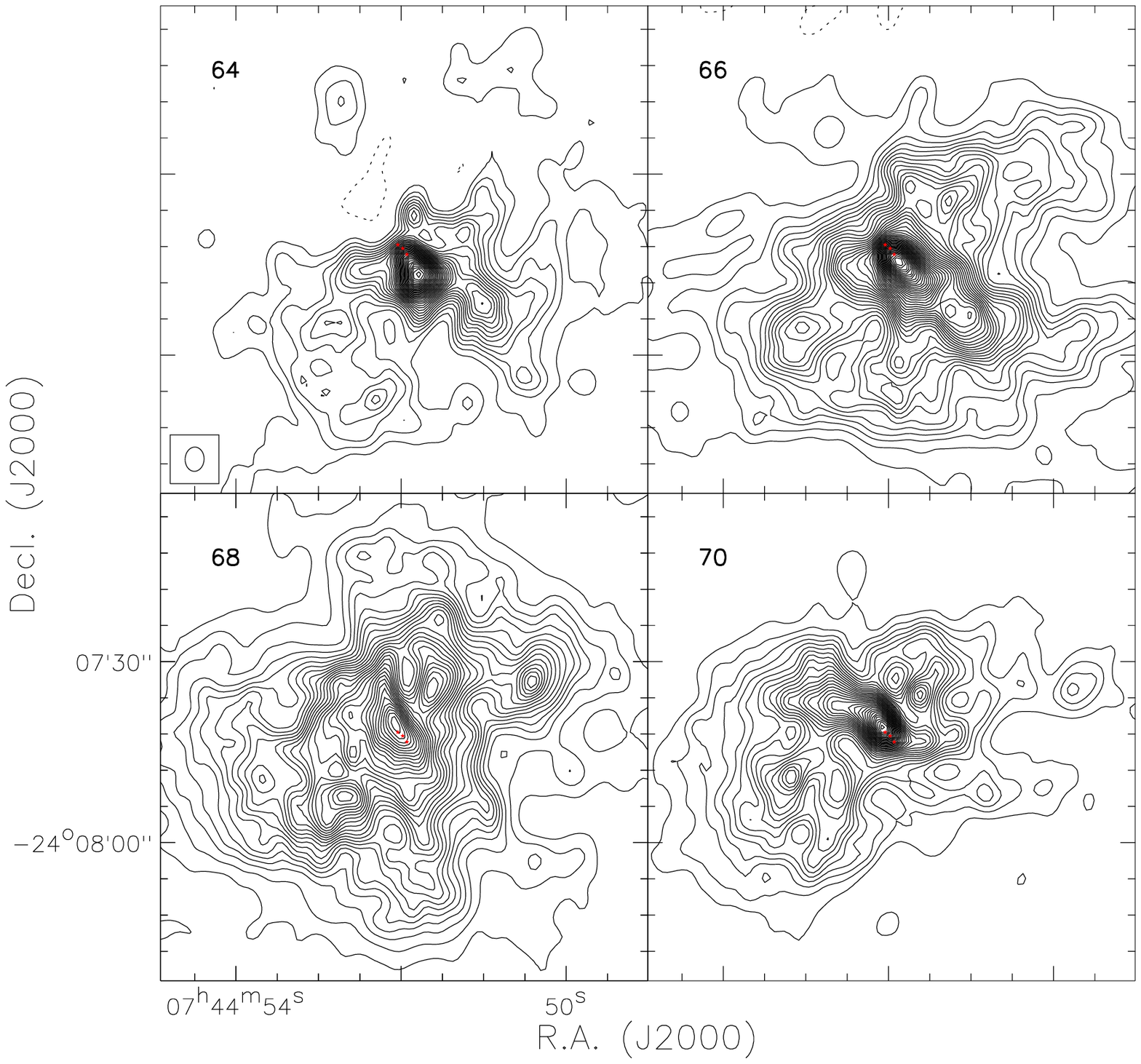} \caption{Combined SMA and CSO
$^{13}$CO (2--1) channel maps of 64--70 km s$^{-1}$, with
solid/dashed contours starting from and continuing at steps of
0.3/-0.3 Jy. \label{smacso13ch}}
\end{figure}

Channel maps of the $^{13}$CO emission are shown in Figures
\ref{sma13ch} and \ref{smacso13ch}, for the SMA alone and the
combined SMA and CSO data around $v_{cloud}$, respectively.
Channels of 64--70 km s$^{-1}$ are dominated by surrounding
diffuse gas but also show extension along the outflow orientation.
There is an embedded dense clump showing a northeast-southwest
velocity gradient, which is also detected in other dense gas
tracers. Emission from the bipolar outflow dominates channel of
${\leq}$62 km s$^{-1}$ and ${\geq}$72 km s$^{-1}$. The $^{13}$CO
emission appears confined within 10 km s$^{-1}$ from $v_{cloud}$,
for both blue- and redshifted lobes.

\subsubsection{Outflow mass and energetics} We use the combined SMA
and CSO data to estimate the gas mass of the outflow, following
$$M_{out}=1.39\times10^{-6}{\mathrm {exp}}(\frac{16.59}{T_{ex}})(T_{ex}+0.92)D^2\int\frac{{\tau}_{12}}{1-e^{-{\tau}_{12}}}S_{\nu}\,dv\,,$$
where $M_{out}$, $T_{ex}$, $D$, ${\tau}_{12}$, and $S_{\nu}$ are
the outflow gas mass in M$_{\odot}$, excitation temperature of the
$^{12}$CO (2--1) transition, source distance in kpc, optical depth
for $^{12}$CO (2--1), and line flux in Jy, respectively. We adopt
an excitation temperature of 30 K, a $^{12}$CO-to-H$_2$ abundance
of $10^{-4}$, and a mean gas atomic weight of 1.36. Assuming the
same excitation temperature for $^{12}$CO (2--1) and $^{13}$CO
(2--1), we calculate ${\tau}_{12}$ according to
$$\frac{S_{\nu}(^{12}\mathrm{CO})}{S_{\nu}(^{13}\mathrm{CO})}=\frac{1-e^{-{\tau}_{12}}}{1-e^{-{\tau}_{12}/{\chi}}}\,,$$
where $\chi$ is the abundance ratio of $^{12}$CO to $^{13}$CO,
which is estimated to be 103 at a Galactocentric distance of 12.7
kpc for G240 \citep{Wilson94}. Many $^{12}$CO outflow studies
estimate outflow mass either by assuming that $^{12}$CO emission
is optically thin at all outflow velocities or by correcting for a
constant $^{12}$CO opacity. Recently a few interferometric outflow
studies use single-dish $^{12}$CO/$^{13}$CO line ratios to correct
for $^{12}$CO opacity as a function of velocity \citep{Shepherd98,
Shepherd03, Su04}. Our interferometric, beam-matching observations
of $^{12}$CO and $^{13}$CO emission, both complemented by short
spacing information from single-dish observations, enable a
measurement of spatial and spectral distribution of
$^{12}$CO/$^{13}$CO line ratios, thus a more sophisticated
correction for the $^{12}$CO opacity. We assume the $^{12}$CO
emission is optically thin where the $^{13}$CO is not detected.
The momentum and energy of the outflow is then derived following
$$P_{out}={\Sigma}M_{out}(v)v$$ and $$E_{out}=\frac{1}{2}{\Sigma}M_{out}(v)v^2\,,$$
respectively, where (and hereafter) $v$ denotes the outflow
velocity relative to $v_{cloud}$. To avoid contamination from the
ambient molecular gas, we use channels of $\leq62$ km s$^{-1}$ in
the blueshifted outflow and channels of $\geq72$ km s$^{-1}$ in
the redshifted outflow, although channels of 64 and 70 km s$^{-1}$
show contributions from the outflow (see Figures \ref{sma12ch} and
\ref{smacso12ch}). The area taken into account is defined by a
polygon encompassing the outflow structure shown in Figure
\ref{flow_int}a. The dynamical timescale of the outflow is derived
following $t_{dyn}=L_{flow}/v_{max}$, where $L_{flow}\sim0.65$ pc
and $v_{max}\sim26.5$ km s$^{-1}$ are the half length of the
end-to-end flow extension and the maximum flow velocity relative
to $v_{cloud}$, respectively. Consequently the mass loss rate of
the outflow can be derived following
$\dot{M}_{out}=M_{out}/t_{dyn}$. The derived outflow parameters
are listed in Table \ref{table1}.

\section{Discussion}
\subsection{The outflow powering source}
From Figure \ref{flow_int}, it is evident that one of the three mm
continuum cores harbors the driving source of the bipolar outflow.
From the maser and cm continuum observations in the literature
\citep{Hughes93, Caswell97, Migenes99, Kumar03}, MM1 roughly
coincides with an UC H{\scriptsize II} region and a H$_2$O maser;
MM2 is associated with H$_2$O and OH masers but shows no cm
emission down to 0.5 mJy at 6 cm or 2 mJy at 3.6 cm; MM3 has no
apparent signature of active star formation. The 1.3 mm continuum
emission from MM3 appears much weaker than that from MM1 and MM2,
implying a mass significantly less than 10 M$_\odot$. Its
relatively faint dust emission and the apparent non-association
with active star formation suggest that MM3 does not harbor the
driving source of the bipolar outflow. Of the remaining two mm
cores, MM2 appears closer to the geometric center of the bipolar
outflow, thus is more likely to be the parent core of the outflow
driving source. This is compatible with the relatively short
outflow dynamical timescale, $2.4\times10^4$ yr, in comparison
with $\gtrsim10^5$ yr for massive molecular outflows from UC
H{\scriptsize II} regions \citep{Shepherd98,Shepherd03}.

\subsection{Comparison with low-mass molecular outflows} \label{comparison}
Low-mass molecular outflows are known to exhibit highly collimated
jetlike structures and/or wide-angle quasi-conical or
quasi-parabolic structures. It is still unclear whether massive
molecular outflows have an essentially similar appearance as
low-mass outflows. High-angular-resolution interferometric
observations have resolved a few single-dish bipolar outflows into
multiple jetlike outflows, which were suggested as indirect
evidence that associated high-mass stars may form via disk
accretion \citep[e.g.,][]{Beuther02b, Qiu07, Zhang07a, Zhang07b}.
On the other hand, very few well-defined wide-angle molecular
outflows have been found in high-mass star-forming regions.
\citet{Qiu08} reported bipolar conical outflow cavities in {\it
Spitzer} mid-IR imaging of IRAS 20126+4104 and HH 80/81, but the
former is subject to ambiguity about whether it is caused by a
precessing jet and the latter lacks kinematical information due to
the lack of high-angular-resolution mm line observations. The
G192.16-3.82 outflow, originating from an UC H{\scriptsize II}
region, may be representative of a bipolar wide-angle outflow with
limb-brightened shells in a $10^3$ L$_{\odot}$ star forming region
\citep{Shepherd98}. Our observations of the G240 outflow reveal a
bipolar wide-angle molecular outflow with a quasi-parabolic shell.
To our knowledge this is hitherto the best case of a well-defined
wide-angle molecular outflow in a $>10^4$ L$_{\odot}$ star forming
region. The outflow has a morphology similar to protostellar
low-mass molecular outflows, e.g., L1228 outflow \citep{Arce04},
B335 outflow \citep{Hirano88, Jorgensen07}. The behavior of higher
collimation at higher velocities is also often seen in low-mass
molecular outflows.

Given the striking similarity between the G240 outflow and
low-mass wide-angle outflows, the G240 outflow could be driven by
an underlying wide-angle wind, as commonly suggested for low-mass
wide-angle outflows. \citet{Churchwell97} argued that neither
accumulated stellar winds nor entrained interstellar matter in
bipolar jets is able to account for $\gtrsim10$ M$_{\odot}$ of gas
masses measured in massive molecular outflows typically at
$\sim10^4$ yrs old. It is true that the G240 molecular outflow is
unlikely to primarily contain accumulated stellar winds, because
stellar winds from OB stars are radiation driven and their
momentum supply rates cannot be higher than $L_{bol}/c$, while the
G240 outflow has a mechanical force, $P_{out}/t_{dyn}$, about 25
times higher than $L_{bol}/c$. The mass limit of entrained
interstellar matter argued by \citet{Churchwell97} was based on
the turbulent jet entrainment model \citep{Canto91}. In this model
a bipolar jet entrains aside interstellar matter through a
turbulent mixing layer. Because the environment cannot react to
the rarefaction by the entrainment at velocities faster than the
sound speed $c_0$ (typically an order of 1 km s$^{-1}$), the
entrainment rate per unit area is given by
${\epsilon}{\rho}_0c_0$, where ${\epsilon}\ll1$ is the entrainment
efficiency and ${\rho}_0$ the density of the ambient medium. On
the other hand, \citet{Shu91} proposed an analytical model for
bipolar molecular outflows, where an underlying wide-angle wind
radially sweeps up ambient gas into an expanding shell to form a
molecular outflow. Further analytical calculations and simulations
of the model, by adopting an angle-dependent wind blowing into an
angle-dependent molecular cloud core, appear generally consistent
with observations \citep{Li96, Lee01, Shang06}. In this model
molecular outflows are assumed to be {\it momentum driven} and the
entrainment rate per unit area is simply proportional to
${\rho}_0v_s$, where $v_s$ is the shell expanding speed (i.e., the
molecular outflow speed observed in $^{12}$CO, typically an order
of 10 km s$^{-1}$) and determined by the momentum supply rate of
the underlying wind. The mass loading limit in the turbulent
entrainment model does not seem to be applicable to the swept-up
shell model. Even in the turbulent entrainment model, a factor of
10 higher mass loading could be achieved if the entrainment
process works at Alfven speeds, which are a factor of 10 higher
than $c_0$, when the magnetic fields are important. Another
molecular outflow model that can readily account for high mass
loading is the so-called ``circulation model''
\citep{Churchwell97, Lery02}. Molecular outflow materials in this
model are those deflected from infalling matter, instead of
entrained or swept-up ambient gas by an underlying jet or wind.
However, in at least one case, the G192.16 outflow, there is
strong evidence for the presence of a wide-angle wind being the
driving agent of the wide-angle $^{12}$CO outflow \citep[][and
references therein]{Qiu08}. The 3.6 and 4.5 $\mu$m emission
coincident with the G240 $^{12}$CO outflow lobes shown in Figure
\ref{flow_int}d, is likely excited by an underlying fast wind.
Therefore, we suggest that the G240 outflow is more likely to be
driven/swept-up by an underlying wide-angle wind. The wind itself
might be accretion driven and somewhat analogous to {\it X}-wind
or disk-wind in low-mass stars \citep{Shu00, Pudritz07}, although
it is unknown whether {\it X}-wind or disk-wind can be scaled up
and applied to high-mass stars.

\begin{figure}
\plotone{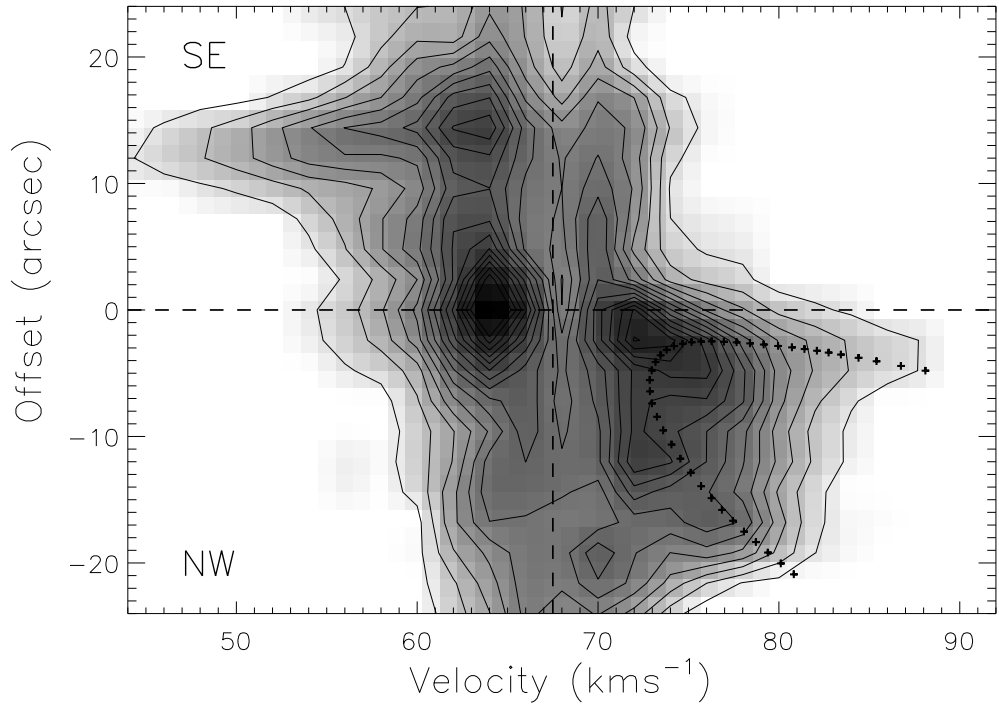} \caption{Position-velocity diagram of the
$^{12}$CO outflow, cut along a position shown as a solid line in
Figure \ref{flow_int}b. The contour levels go from 1 to 22 Jy
beam$^{-1}$ in steps of 1.5 Jy beam$^{-1}$, and the gray scale
stretches in a square-root algorithm from 0.5 Jy beam$^{-1}$ ({\it
white}) to the peak emission at 27.8 Jy beam$^{-1}$ ({\it black}).
Plus signs outlines a parabolic trajectory (see the discussion in
\S \ref{comparison} for details). The vertical dashed line
indicates $v_{cloud}$ and the horizontal dashed line the position
of the potential driving source of the outflow, i.e, the MM2 peak.
The diagram is constructed from the combined SMA and CSO data.
\label{pv}}
\end{figure}

Figure \ref{pv} shows the position-velocity (PV) diagram of the
G240 $^{12}$CO outflow cut roughly along the outflow major axis.
The NW redshifted lobe shows a feature extending to $\sim5''$ from
the driving source at a velocity of $\sim88$ km s$^{-1}$ and
another feature extending to $\sim20''$ from the source at a
velocity of $\sim80$ km s$^{-1}$. Connecting these two features
following the gradient of the contour levels in Figure \ref{pv}
one can get an approximately parabolic trajectory opening outward
from the driving source. This kind of PV structure is very similar
to that of the northern lobe of the VLA 05487 outflow, a good
example of a wide-angle wind driven low-mass outflow, and can be
reasonably reproduced by the hydrodynamic simulation of a modified
{\it X}-wind model \citep{Lee01}. According to the model the PV
structure can be qualitatively understood in the context of a
radially expanding shell with a Hubble-law velocity structure and
a moderate inclination angle, where the far-side shell has a
smaller projected extension and higher line-of-sight velocities,
while the near-side shell has a larger projected extension and
lower line-of-sight velocities. This parabolic PV structure
supports the scenario that the G240 molecular outflow is driven by
a wide-angle wind. It is difficult to identify a similar PV
structure in the SE blueshifted lobe, probably because this lobe
is highly asymmetric at most velocities (the western part of the
shell can be only seen at 62 and 64 km s$^{-1}$ in Figure
\ref{sma12ch}) and contaminated by a secondary outflow to the
west. Its PV structure is dominated by a high velocity feature
$\sim14''$ away from the source, corresponding to a clump shown in
Figure \ref{flow_int}b.

\begin{figure}
\plotone{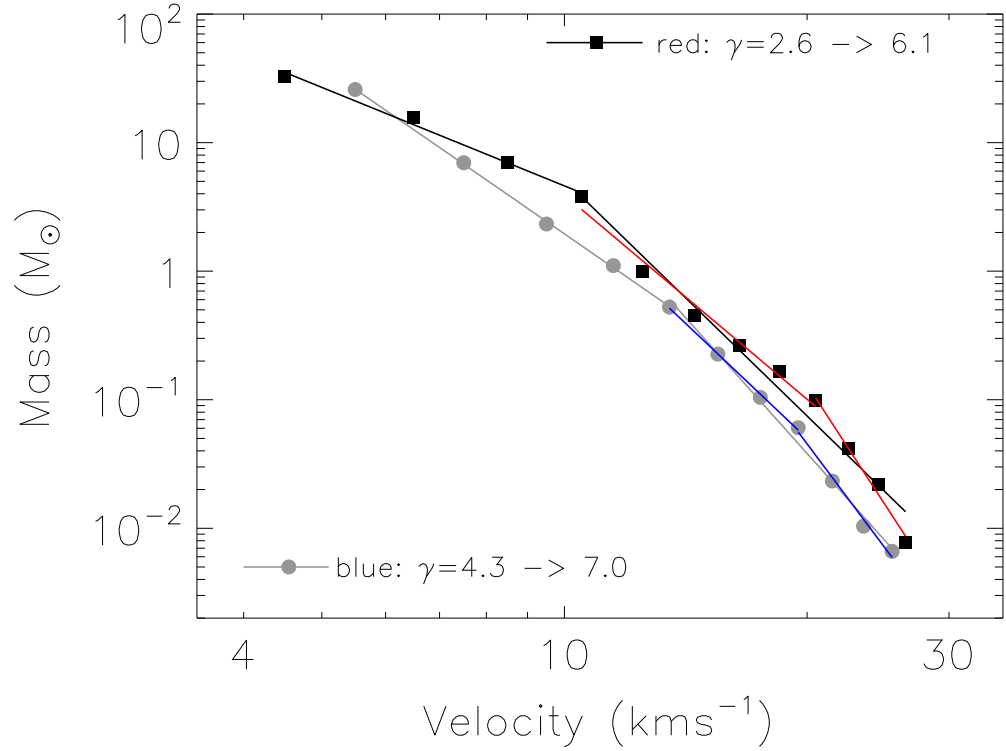} \caption{Mass--velocity diagram of the $^{12}$CO
outflow, calculated from the combined SMA and CSO data for the gas
mass in each 2 km s$^{-1}$ wide channel. Black filled squares
denote the measurements from the blueshifted lobe, and gray filled
circles the redshifted lobe. Solid black and gray lines indicate
broken power-law fittings to the blue- and redshifted lobes,
respectively; the blue and red lines represent further broken
power-law fittings in the specific velocity ranges, rather than a
single power-law fitting. \label{mv}}
\end{figure}

While molecular outflows associated with both low- and high-mass
(proto)stars are found to exhibit a ``mass spectrum'',
$m(v)\,{\propto}\,v^{-\gamma}$, with the power-law index $\gamma$
around 2 for low velocity gas and often steepening at velocities
higher than $\sim10$ km s$^{-1}$, its physical origin is not well
established \citep{Richer00, Arce07}. Figure \ref{mv} shows the
mass-velocity diagram of the G240 outflow, where both lobes can be
fitted as a broken power-law, with $\gamma$ steepening from
$4.3\pm0.1$ to $7.0\pm0.3$ at 13.5 km s$^{-1}$ for the blueshifted
lobe and from $2.6\pm0.2$ to $6.1\pm0.4$ at 10.5 km s$^{-1}$ for
the redshifted lobe. It appears that the G240 outflow has fairly
large values of $\gamma$ at high velocities, comparable to that of
some other $^{12}$CO outflows associated with both high- and
low-mass (proto)stars \citep{Shepherd98, Yu00, Arce01, Su04}. Such
large values of $\gamma$ have been an unsolved issue to both jet
bow-shock and wide-angle wind models since numerical simulations
of these models predict $\gamma<4$; future simulations that run
for flows more than $10^4$ yr old might be able to address whether
it can be explained by outflow evolution \citep[see a review
by][]{Arce07}. Moreover, as shown by colored lines in Figure
\ref{mv}, the mass spectra of high velocity channels of both lobes
can be further fitted with a broken power-law rather than a single
power-law, and in those narrower velocity ranges $\gamma$ steepens
from $5.9\pm0.2$ to $8.4\pm0.7$ at 19.5 km s$^{-1}$ for the
blueshifted lobe and from $5.3\pm0.3$ to $9.6\pm0.8$ at 20.5 km
s$^{-1}$ for the redshifted lobe. This may simply reflect a
continuing decrease of mass entrainment efficiency with increasing
outflow velocities in this outflow.

\subsection{Mass and energetics}
The total gas mass of the G240 outflow, estimated based on
combined interferometer and single-dish data with sophisticated
opacity correction, amounts to 98 M$_{\odot}$, making it one of
the most massive (proto)stellar outflows known. Given the unknown
inclination angle of the outflow, the estimated outflow momentum
of 620 M$_{\odot}$ km s$^{-1}$ and energy of $4.5\times10^{46}$
erg can be treated as lower limits accordingly. Based on a
velocity gradient in their CSO C$^{18}$O (2--1) observations and
assuming the gas is in virial equilibrium, \citet{Kumar03} derived
a gas mass of 625 M$_{\odot}$. Considering the cloud is likely
infalling to feed the accretion onto the central (proto)stars and
thus deviates from virial equilibrium, this estimate may provide a
lower limit of the dense cloud mass. The gravitational binding
energy of the cloud, $GM^2_{cloud}/R_{cloud}$, where
$R_{cloud}=0.63$ pc \citep{Kumar03}, is then
$\gtrsim4.4\times10^{46}$ erg, which is comparable to the outflow
energy (Table \ref{table1}). Although there are many uncertainties
in calculations of both the gravitational binding energy and the
outflow energy, the G240 outflow seems to play a critical role in
disrupting or unbinding its host cloud. The mass loss rate of
$3.8\times10^{-3}$ M$_{\odot}$ yr$^{-1}$ is about two to three
orders of magnitude higher than that of typical low-mass outflows.
Assuming momentum conservation between the swept-up molecular
outflow and an underlying primary wind, i.e.,
$P_{out}/t_{dyn}=\dot{M}_wv_w$ \citep[also see Eq. 8 of][]{Shu91},
where $\dot{M}_w$, $v_w$ are the wind mass loss rate and wind
speed, respectively, and adopting a wind speed of 500 km s$^{-1}$,
the mass loss rate of the wind amounts to $\sim0.5\times10^{-4}$
M$_{\odot}$ yr$^{-1}$. Furthermore, adopting a ratio of
$\frac{1}{3}$ for the wind mass loss rate to mass accretion rate
\citep{Tomisaka98, Shu00}, we get an accretion rate of
$1.5\times10^{-4}$ M$_{\odot}$ yr$^{-1}$, high enough to overcome
radiation pressure from the central object to form a high-mass
star \citep{Wolfire87, Jijina96, Yorke02}. Again considering the
underestimate to $P_{out}$, this estimate may represent a lower
limit to the actual accretion rate.

\section{Summary}
We have presented a high-angular-resolution study of the molecular
outflow in a $10^{4.7}$ L$_{\odot}$ star forming region, using the
$^{12}$CO and $^{13}$CO (2--1) observations from the SMA and CSO,
as well as the {\it Spitzer} IRAC color-composite image. The SMA
1.3 mm dust continuum images are used to pinpoint the outflow
powering source.

A bipolar, wide-angle, quasi-parabolic molecular outflow is
revealed in the $^{12}$CO and $^{13}$CO emission at velocities
more than 3 km s$^{-1}$ apart from the cloud velocity. The outflow
is most likely powered by a deeply embedded high-mass protostar
prior to forming a significant UC H{\scriptsize II} region. From
its morphological similarity to some low-mass molecular outflows
and parabolic PV structure in the redshifted lobe, the G240
outflow is likely driven/swept-up by an underlying wide-angle
wind. For both lobes the outflow mass-velocity relation can be
fitted with broken power-laws with the indices steeping at high
velocities. The outflow has $\sim98$ M$_{\odot}$ gas mass and a
dynamical timescale of $2.4\times10^4$ yr, thus a mass loss rate
of $4.1\times10^{-3}$ M$_{\odot}$ yr$^{-1}$. Without correction
for the unknown inclination angle the momentum and energy of the
outflow amount to 620 M$_{\odot}$ km s$^{-1}$ and
$4.5\times10^{46}$ erg, respectively.

Our observations of the G240 outflow reveal a well-defined
wide-angle molecular outflow with its morphology and kinematics
similar to low-mass molecular outflows, but with two to three
orders of magnitude greater mass and energetics suggestive of a
high accretion rate. These findings appear to be consistent with
the scenario that high-mass stars of early-B to late-O types can
form in a similar way as low-mass star formation, even in a
clustering environment.

\acknowledgments This work is based in part on observations made
with the {\it Spitzer Space Telescope}, which is operated by the
Jet Propulsion Laboratory, California Institute of Technology
under a contract with the National Aeronautics and Space
Administration (NASA).

\clearpage

\clearpage

\clearpage

\clearpage

\clearpage

\clearpage

\clearpage

\clearpage

\clearpage

\clearpage

\begin{deluxetable}{cccccccccc} \rotate
\tablewidth{0pc} \tablecaption{Derived core and outflow parameters
\label{table1}} \tablehead{ \colhead{source} &
\colhead{$M_{env}$\tablenotemark{1}} &
\colhead{$M_{core}$\tablenotemark{2}} &
\colhead{$M_{blue}$\tablenotemark{3}} &
\colhead{$M_{red}$\tablenotemark{4}} & \colhead{$M_{out}$} &
\colhead{$P_{out}$} & \colhead{$E_{out}$} &
\colhead{$t_{dyn}$} & \colhead{$\dot{M}_{out}$} \\
\colhead{} & \colhead{(M$_{\odot}$)} & \colhead{(M$_{\odot}$)} &
\colhead{(M$_{\odot}$)} & \colhead{(M$_{\odot}$)} &
\colhead{(M$_{\odot}$)} & \colhead{(M$_{\odot}$ km s$^{-1}$)} &
\colhead{($10^{46}$erg)} & \colhead{($10^4$yr)} &
\colhead{($10^{-3}$ M$_{\odot}$yr$^{-1}$)} }
\startdata  G240.31+0.07  & 150  & 40 & 37 & 61 & 98 & 620 & 4.5 & 2.4 & 4.1  \\
\enddata
\tablenotetext{1}{Mass of the dusty envelope as shown in Figure
\ref{continuum}a} \tablenotetext{2}{Combined mass of the three
dusty cores as shown in Figure \ref{continuum}b} \tablenotetext{3}
{Outflow mass in the blueshifted lobe} \tablenotetext{4}{Outflow
mass in the redshifted lobe}
\end{deluxetable}

\end{document}